\documentclass[preprint,showpacs,showkeys,preprintnumbers,aps]{revtex4}

\usepackage{graphicx}
\usepackage{float}

\begin{document}

\preprint{Investigation of Co$_2$FeSi, Wurmehl \it{et al}}

\title{Investigation of Co$_2$FeSi: The Heusler compound
       with Highest Curie Temperature and Magnetic Moment.}

\author{Sabine~Wurmehl, Gerhard~H.~Fecher, Hem~Chandra~Kandpal,
        Vadim~Ksenofontov, and Claudia~Felser}
\email{felser@uni-mainz.de}
\affiliation{
Institut f\"ur Anorganische und Analytische Chemie,
Johannes Gutenberg - Universit\"at, 55099 Mainz, Germany.}

\author{Hong-Ji~Lin}
\affiliation{National Synchrotron Radiation Research Center - NSRRC,
Hsinchu, 30076, Taiwan.}

\date{\today}

\begin{abstract}
This work reports on structural and magnetic investigations of the
Heusler compound Co$_2$FeSi. X-Ray diffraction and M\"o\ss bauer
spectrometry indicate an ordered $L2_1$ structure. Magnetic
measurements by means of X-ray magnetic circular dichroism and
magnetometry revealed that this compound is, currently, the material
with the highest magnetic moment ($6 \mu_B$) and Curie-temperature
(1100K) in the classes of Heusler compounds as well as half-metallic
ferromagnets.
\end{abstract}



\maketitle

Materials that exhibit half-metallic ferromagnetism are seen to be
potential candidates for the field of application being called
spintronics \cite{CVB02, ZFS04}, that is electronics making use of
electron spin instead of its charge. The concept of
half-metallic ferromagnetism was first presented by de Groot
\cite{GME83}, predicting it to appear in half Heusler compounds.
The model suggests that the density of states exhibits, around the
Fermi energy ($\epsilon_F$), a gap for minority electrons. Thus,
these materials are supposed to be 100\% spin polarized at
$\epsilon_F$. Most of the predicted half-metallic ferromagnets
(HMF) belong to the Heusler \cite{Heu03} compounds. In general,
these are ternary $X_2YZ$-compounds crystallizing in the $L2_1$
structure. $X$ and $Y$ are usually transition metals and Z is a
main group element.

The Co$_2$ based Heusler compounds exhibit the highest Curie
temperature ($T_C=985$K, Co$_2$MnSi \cite{BEJ83}) and the highest
magnetic moment ($5.54 \mu_B$ per unit cell, Co$_2$FeGe
\cite{BEJ83}) being reported up to now. High Curie temperatures,
magnetic moments, and large minority gaps are desirable for
applications. For room temperature devices, in particular, one needs
to prevent a reduction of the magnetic properties by thermal
effects. The HMF character of Co$_2$MnZ compounds was first reported
by Ishida et al. \cite{IFK95}. Recently, Co$_2$MnSi \cite{KTH04} and
Co$_2$MnGe \cite{DAX05} were used to built first thin film devices.
The present work reports on structural and magnetic properties of
Co$_2$FeSi.

In the past, this compound was reported to exhibit a magnetic moment
of only $5.18 \mu_B$ per unit cell and a Curie temperature of above
$980K$ \cite{NBR77, Bus88}. One expects, however, that it has a
magnetic moment of $6\mu_B$ if following the rule of thumb
($m=N_V-24$, where $N_V$ is the number of accumulated valence
electrons in the unit cell) based on the Slater Pauling rule
\cite{Kue84,GDP02}.

Co$_2$FeSi samples were prepared by arc-melting of stochiometric
quantities of the constituents in an argon atmosphere. Care was
taken to avoid oxygen contamination. Afterwards, the polycrystalline
ingots were annealed in an evacuated quartz tube at $1300K$ for 20
days. This procedure resulted in samples exhibiting the correct
Heusler type structure ($L2_1$) as was proved by X-ray diffraction
(XRD) using Cu-K$_\alpha$ and Mo-K$_\alpha$ radiation. The lattice
constant was determined to be {5.64\AA} from Rietveld-refinement.
The $R_{Bragg}$-value was estimated to be $<5.5$. A disorder between
Co and Fe atoms ($DO_3$ type disorder) can be excluded from the XRD
Rietveld refinement (see Fig.1) as well as from Neutron scattering
data (not shown here). A small disorder between Fe and Si ($B2$ type
disorder) atoms ($<10$\% \footnote{A $B2$ like disorder of below 0.1
does not effect the spin polarization at the Fermi-energy
\cite{MNS04} because this type of disorder does not perturb the
simple cubic Co sublattice restricting the gap in the minority
states \cite{FKW05}.}) can not be excluded from neither of these
methods, particularly due to the low intensities of the (111) and
(200) diffraction peaks in XRD. The lattice parameter is obviously
smaller than reported \cite{Bus88} and a lower degree of disorder is
observed in the present work (compare to Ref.:\cite{NBH79}).

\begin{figure}
\centering
\includegraphics[width=8.5cm]{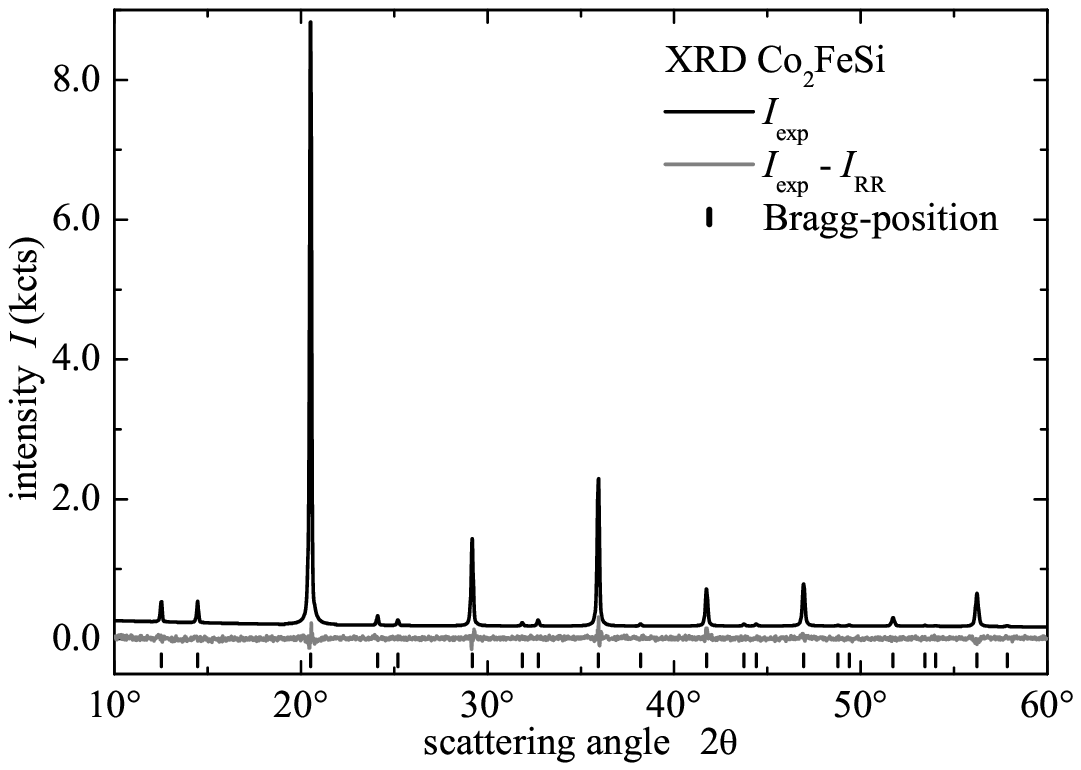}
\caption{XRD of Co$_2$FeSi. \\
         Shown are the measured intensity
         ($I_{exp}$) and the difference to the Rietveldt refinement
         ($I_{RR}$). Vertical bars indicate the Bragg positions.}
\label{fig_1}
\end{figure}

For further structural and magnetic investigations, M\"o\ss bauer
spectroscopy was performed in transmission geometry using a constant
acceleration spectrometer with a source line width of 0.105mm/s
($^{57}$Co(Rh)). The observed $^{57}$Fe M\"o\ss bauer line width of
0.15mm/s is characteristic for a well-ordered system. The value is
comparable to 0.136mm/s observed from $\alpha$-Fe at 4.2K. In
detail, the M\"o\ss bauer spectrum exhibits a sextett with an isomer
shift of 0.23mm/s and a hyperfine magnetic field of $26.3 \times
10^6$A/m at 85K. No quadrupole splitting was detected as expected
for the cubic symmetry of the local Fe environment. A $DO_3$ like
disorder can be definitely excluded by comparing measured and
calculated hyperfine fields in ordered and disordered structures.

\begin{figure}
\centering
\includegraphics[width=8.5cm]{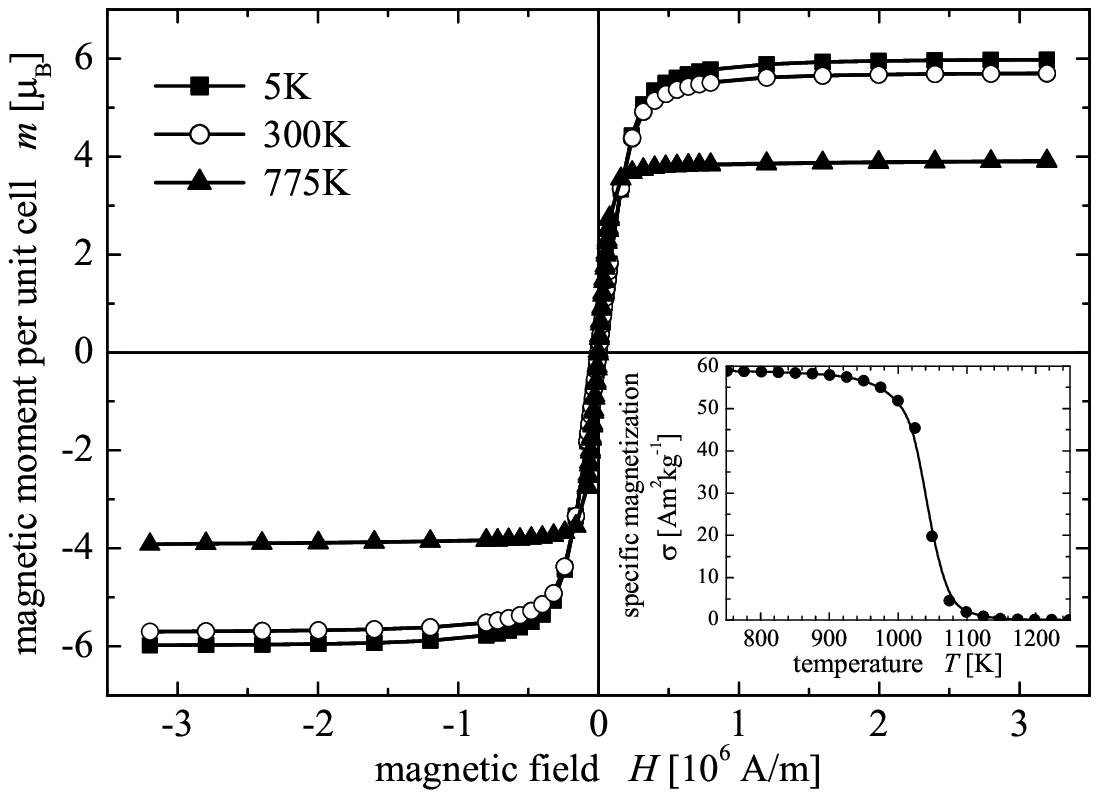}
\caption{Magnetic properties of Co$_2$FeSi. \\
         The field dependence of the magnetic moments was measured by SQUID
         magnetometry at different temperatures. The inset shows the temperature
         dependence of the specific magnetization measured by VSM.}
\label{fig_2}
\end{figure}

Low temperature magnetometry was performed using a super conducting
quantum interference device (SQUID) to proof the estimated
saturation moment. The results are shown in Fig.\ref{fig_2}. The
measured magnetic moment in saturation is $(5.97 \pm 0.05)\mu_B$ at
5K corresponding to $1.49\mu_B$ per atom. This value is obviously
larger than the previously reported ($5.18\mu_B$ see: \cite{NBH79,
Bus88}). An extrapolation to $6\mu_B$ per unit cell at 0K fits
perfectly to the moment estimated from the Slater-Pauling rule \cite
{KWS83, Kue84, GDP02}. The measurement of the magnetic moment
reveals, as expected for a HMF, an integer within the experimental
uncertainty. Regarding the result of the measurement (an integer)
and the rule of thumb, it all sums up to an evidence for
half-metallic ferromagnetism in Co$_2$FeSi. In more detail,
Co$_2$FeSi turns out to be soft magnetic with a small remanence of
$\approx0.3$\% of the saturation moment and a small coercive field
of $\approx750$A/m, under the experimental conditions used here.

The experimental magnetic moment is supported by band structure
calculations revealing a HMF character with a magnetic moment of
$6\mu_B$. The results of these calculations will be given elsewhere
\cite{Hem}.

\begin{figure}
\centering
\includegraphics[width=8.5cm]{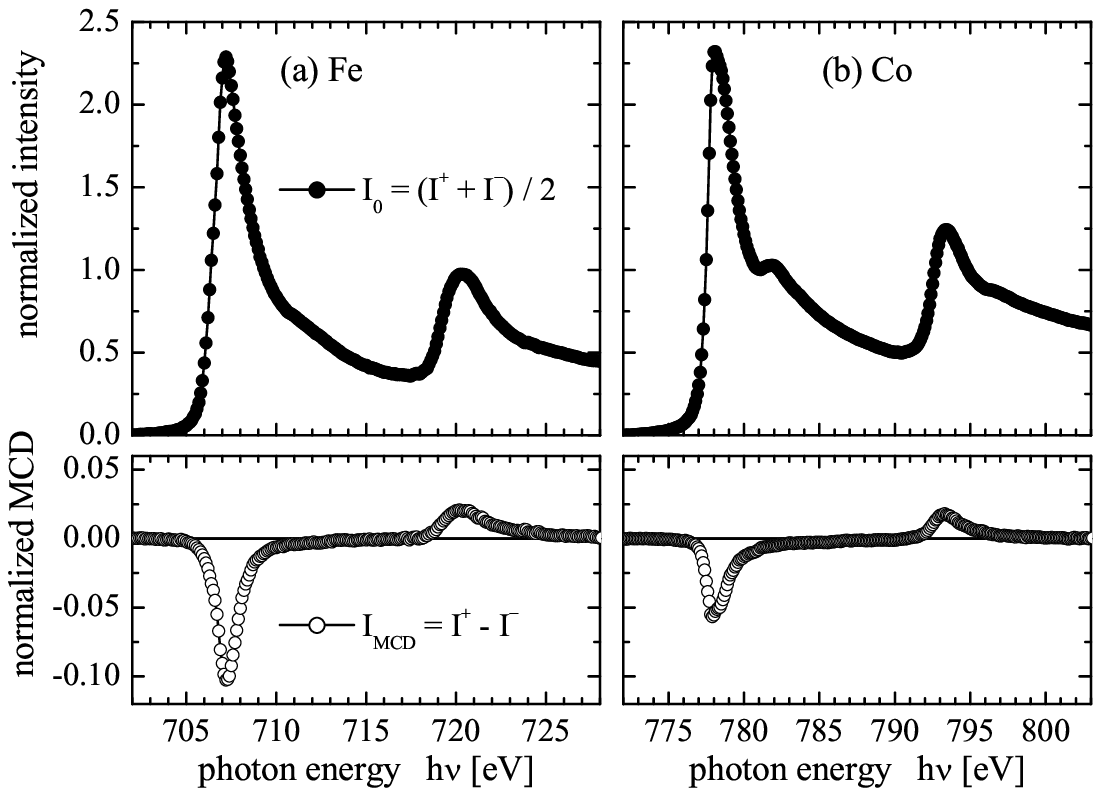}
\caption{Site resolved magnetic properties of Co$_2$FeSi. \\
         Shown are the XAS (I$_0$) and XMCD (I$_{MCD}$) spectra taken at the $L_{2,3}$
         absorption edges of Fe (a) and Co (b) after subtracting a constant background.}
\label{fig_3}
\end{figure}

X-ray magnetic circular dichroism (XMCD) in photo absorption (XAS)
was measured at the {\it First Dragon} beamline of NSRRC (Hsinchu,
Taiwan). The XAS and XMCD spectra taken at the $L_{2,3}$ absorption
edges of Fe and Co are shown in Fig.\ref{fig_3}. The feature seen at
3eV below the $L_3$ absorption edge of Co is related to the $L2_1$
structure and points on the high structural order of the sample (it
vanishes for $B2$ like disorder). The magnetic moments per atom
derived from a sum rule analysis \cite{TCS92,CTA93} are
$(2.6\pm0.1)\mu_B$ for Fe and $(1.2\pm0.1)\mu_B$ for Co, at 300K
and $\mu_0H=0.4$T. The error arises mainly from the unknown number
of holes in the $3d$ shell and the disregard of the magnetic dipole
term in the sum rule analysis. The orbital to spin magnetic moment
ratios are about 0.05 for Fe and 0.1 for Co. All values are in good
agreement with electronic structure calculations \cite{Hem}.

Inspecting the magnetic data of the known Heusler compounds (see
data and references in \cite{LB19C,LB32C}), one finds a very
interesting aspect. Seemingly, a linear dependence (not shown here)
is obtained for Co$_2$ based Heusler compounds when plotting the
Curie temperature ($T_C$) of the known, $3d$ metal based Heusler
compounds as function of their magnetic moment. According to this
plot, $T_C$ is highest for those half-metallic compounds that
exhibit a large magnetic moment, or equivalently for those with a high
valence electron concentration as derived from the Slater-Pauling
rule. $T_C$ is estimated to be above 1000K for compounds with $6\mu_
B$ by an extrapolation from the linear dependence.

The high temperature magnetization of Co$_2$FeSi was measured by
means of a vibrating sample magnetometer (VSM) equipped with a high
temperature stage. The specific magnetization as function of the
temperature is shown in the inset of Fig.\ref{fig_2}. The
measurements were performed in a constant induction field of
$\mu_0H=0.1$T. For this induction field, the specific magnetization
at 300K is 37\% of the value measured in saturation. The
ferromagnetic Curie temperature is found to be $T_C=(1100\pm20)$K.
This value fits very well to the linear behavior of $T_C$ as a
function of valence electrons for Co$_2$ based Heusler compounds as
mentioned above. $T_C$ is well below the melting point being
obtained by means of differential scanning calorimetry to be
$T_m=(1520\pm5)$K.

The highest known Curie temperature is reported for elemental Co to
be $1388$K \cite{CRC01}. Only few materials exhibit a $T_C$ above
$1000$K, for example the Fe-Co binary alloys. With a value of
$\approx1100$K, Co$_2$FeSi has a higher Curie temperature than Fe
and the highest of all HMF and Heusler compounds being measured up
to now.

In summary, the present work shows that $L2_1$ ordered Co$_2$FeSi is
a half-metallic ferromagnet exhibiting the highest values for Curie
temperature and magnetic moment reported for full Heusler compounds.

\bigskip
\begin{acknowledgments}
We thank Y.~Hwu (Academia Sinica, Taipei, Taiwan) and F.~Casper
(Mainz) for help with the experiments. We acknowledge assistance from
Lake Shore Cryotronics, Inc. in the high
temperature magnetic measurements using the VSM Model 7300 system
equipped with a high temperature oven Model 73034. Further, we
thank G.~Frisch (Ludwig Alberts-University, Freiburg) for
performing Mo-K$_\alpha$ X-Ray diffraction.\\
This work is financially supported by DFG (research project FG 559)
and DAAD (D/03/31 4973).
\end{acknowledgments}

\end{document}